\documentclass[journal]{IEEEtran}
\usepackage{amsmath,amsfonts}
\usepackage{algorithmic}
\usepackage{algorithm}
\usepackage{array}
\usepackage[caption=false,font=normalsize,labelfont=sf,textfont=sf]{subfig}
\usepackage{textcomp}
\usepackage{stfloats}
\usepackage{url}
\usepackage{verbatim}
\usepackage{graphicx}
\usepackage{cite}
\usepackage{amsmath,amssymb,amsfonts}
\usepackage{xcolor}
\usepackage{booktabs}
\usepackage{pifont}
\usepackage{color}
\usepackage{multirow}
\usepackage{makecell}
\begin{document}

\title{MAD-MulW: A Multi-Window Anomaly Detection Framework for BGP Security Events}

\author{Songtao~Peng,
        Yiping~Chen,
        Xincheng~Shu,
        Wu~Shuai,
        Shenhao~Fang,
        Zhongyuan~Ruan,
        and Qi~Xuan,~\IEEEmembership{Senior~Member,~IEEE}
\thanks{S. Peng, Y. Chen, X. Shu, W. Shuai, S. Fang, Z. Ruan, and Q. Xuan are with the Institute of Cyberspace Security, College of Information Engineering, Zhejiang University of Technology, Hangzhou 310023, China. (Corresponding author: Zhongyuan~Ruan, e-mail: zyruan@zjut.edu.cn.)

S. Peng, Z. Ruan, and Q. Xuan are also with Binjiang Cyberspace Security Institute of ZJUT, Hangzhou 310056, China.
}
}

\markboth{Journal of \LaTeX\ Class Files,~Vol.~10, No.~8, September~2023}%
{Peng \MakeLowercase{\textit{et al.}}: MAD-MulW: A Multi-Window Anomaly Detection Framework for BGP Security Events}


\maketitle

\begin{abstract}
 In recent years, various international security events have occurred frequently and interacted between real society and cyberspace. Traditional traffic monitoring mainly focuses on the local anomalous status of events due to a large amount of data. BGP-based event monitoring makes it possible to perform differential analysis of international events. For many existing traffic anomaly detection methods, we have observed that the window-based noise reduction strategy effectively improves the success rate of time series anomaly detection. Motivated by this observation, we propose an unsupervised anomaly detection model, MAD-MulW, which incorporates a multi-window serial framework. Firstly, we design the W-GAT module to adaptively update the sample weights within the window and retain the updated information of the trailing sample, which not only reduces the outlier samples' noise but also avoids the space consumption of data scale expansion. Then, the W-LAT module based on predictive reconstruction both captures the trend of sample fluctuations over a certain period of time and increases the interclass variation through the reconstruction of the predictive sample. Our model has been experimentally validated on multiple BGP anomalous events with an average F1 score of over 90\%, which demonstrates the significant improvement effect of the stage windows and adaptive strategy on the efficiency and stability of the timing model.
\end{abstract}

\begin{IEEEkeywords}
Anomaly Detection, Time Series, Unsupervised Model, Multi-Window
\end{IEEEkeywords}

\section{Introduction}
\IEEEPARstart{T}{oday}, the rapid development of the Internet has provided quality services in the fields of business, education, and entertainment. With society's increasing dependence on the Internet, its reliability and security are of great concern. Researchers have been devoting themselves to discovering and alerting various deliberate attacks, such as distributed denial-of-service attacks, network worms, IP hijacking, and other attack threats. At the same time, international conflicts such as the Russia-Ukraine war or natural disasters such as earthquakes and typhoons can also cause widespread network fluctuations or even breakdowns due to network cascading, causing incalculable damage. Therefore, the detection and early alert of endless anomalies are essential to secure the network. In large-scale anomalous events, traffic monitoring at the interface level is limited to a narrow view. As the most widely used inter-domain routing protocol, the Border Gateway Protocol (BGP) makes it possible to detect anomalous events by managing network reachability information.

\begin{figure}[htb]
\centering
\setlength{\abovecaptionskip}{-0.2cm}
{
    \begin{minipage}[b]{\linewidth}
        \centering
        \includegraphics[scale=1.2]{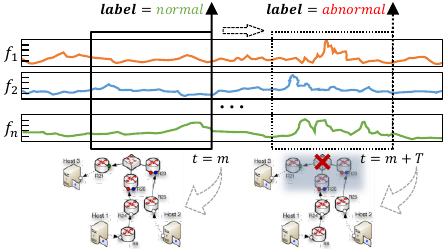}
    \end{minipage}
}
\caption{Multivariate time series in different network connectivity states based on a sliding window perspective (Solid boxes are normal network states and dashed boxes are abnormal network states). }
\label{fig: window_slicing}
\vspace{-3mm}
\end{figure}

BGP-based anomalous events can be defined as an anomaly in which a large number of BGP update messages are generated while part of the communication links are disconnected due to an external force. By parsing, extracting, and counting BGP update packets, one can get the analyzable time series data. In terms of data, Semenoglou et al.\cite{khan2023comparing} summarized the existing data enhancement methods, and showed that the sliding window strategy is useful for most data analysis. Dimoudis et al.\cite{dimoudis2023utilizing} proposed an anomaly detection algorithm for rolling median with an adaptive sliding window to detect anomalous data points in time series. These data enhancement strategies can smooth out the data noise and facilitate the task of classification, prediction, and detection. In terms of models, the field of time series provides many proven theories and techniques. The existing unsupervised methods are mostly based on three categories: Long Short Term Memory (LSTM)-used temporal models, LSTM-AD \cite{thill2019anomaly}, and LSTM-GSVM \cite{ergen2019unsupervised}, which focus on temporal attribute in samples; Autoencoder (AE) structure sequence models, TCN-AE \cite{mo2022unsupervised}, LSTM-AE \cite{hou2019lstm}, build sequence mappings by encoding and decoding; Generative Adversarial Network (GAN) structure adversarial models, MTS-DCGAN \cite{liang2021robust}, MAD-GAN \cite{li2019mad}, generalize the distribution of data. These methods have demonstrated excellent performance in a wide range of fields, including the Internet, industry, and medicine.

Sliding windows are commonly utilized in various models for data enhancement and model design (Figure \ref{fig: window_slicing}). The main purpose is to weaken the effect of anomalous noise points by averaging the values or to extend the sample features by means of scale transformation, and finally to achieve the improvement of detection performance. However, this also brings several unavoidable problems, i.e., there is no fixed optimal window size for different data, and the increase in model space and time resources due to sample expansion. Likewise, when it comes to model construction, windows are often concurrent with LSTM models, focusing on capturing correlations between multiple consecutive samples. This process is difficult to achieve quantitative interpretation and precise application of temporal properties. Therefore, considering the importance of windows in multiple stages of model construction, it is worthwhile to explore methods that are generalizable for multiple data and stable for multiple samples.

Based on the above findings and ideas, we propose a stage-two window for feature remodeling and sequence prediction, respectively. Using anomalous event traffic data as input, we group them through a stage-one window and reshape the features using the GAT module to reflect the state after noise reduction with the tail samples of the window; we use the LAE module to internally predict and externally reconstruct the data through window grouping in the second stage to achieve stable inter-class differences. The overall purpose is to generalize the strong similarity of feature fluctuation trends of normal samples. We take this similarity as a major basis for unsupervised anomaly detection, and the difference in anomalous samples will be expressed as high scores. We validate our model on several Internet traffic anomalies and its performance is outstanding compared to many superior methods. In summary, the contributions of our study are mainly in the following:
\begin{itemize}
\item[$\bullet$] \textbf{Broader perspective:}  Mapping a wide range of anomalous events into time series, focusing not only on security events caused by network attacks but also on the situation of a wide range of network fluctuations caused by accidents, international situations, etc.

\item[$\bullet$] \textbf{More efficient model:} Proposing an unsupervised BGP anomaly detection model under a multi-window perspective, named MAD-MulW contained W-GAT and W-LAE. The W-GAT module achieves noise reduction and enhancement of different data by adaptively optimizing the sample parameters in the window, and the W-LAE module enables differential construction of samples by capturing the reconstruction dependencies of the predicted samples, which is ultimately applicable to the analysis and detection of time series data.

\item[$\bullet$] \textbf{More stable result:} It shows remarkable improvement on several anomalous events and achieves excellent detection performance with a small number of training samples without significant fluctuations with parameter changes.
\end{itemize}

Subsequently, Section II focuses on the related work to anomaly detection. Section III presents a detailed explanation of our model. Section IV contains an introduction to the datasets, comparative models, and evaluation metrics. Section V is an evaluation of the model and an analysis of the experimental results. Finally, a summary of our work is presented in Section VI.

\section{Related Work}
In this section, we describe the BGP anomaly detection work and extend it to the development of anomaly detection techniques for time series in recent years.
\subsection{BGP Anomaly Detection Techniques}
BGP is the Internet's default inter-domain routing protocol used to manage the connectivity between Autonomous Systems (ASes). Over the past 20 years, many BGP events have been captured that threaten the stability of the Internet.

\textbf{Supervised learning-based} BGP anomaly detection has been extensively studied, and some of these methods affect anomaly detection performance mainly by improving the quantity or quality of features. Urbina Cazenave et al. \cite{de2011anomaly} showed that Support Vector Machines (SVM), Decision Trees, and Naive Bayesian methods are evaluated for BGP event classification. Arai et al. \cite{arai2019selection} and Xu et al. \cite{xu2020bgp} extracted features by an importance assessment algorithm to enhance the model efficiency. Innovatively, Sanchez et al. \cite{sanchez2019comparing} introduced graph features to detect BGP anomalies, which are more robust than traditional features. Another class of methods focused on sample temporal relationships. The stacked-LSTM \cite{chauhan2015anomaly} and the MSLSTM \cite{cheng2018multi} both achieved significant improvement in anomaly detection performance by mining temporal information. Peng et al. \cite{peng2022multi} went further by introducing a multi-dimensional attention mechanism to improve performance. The supervised approach requires the labeling of the dataset, so its applicability is limited to specific instances.

\textbf{Unsupervised learning-based} work is more sparse. Earlier statistical-based methods used probabilistic models to detect changes in the data, such as Principal Component Analysis (PCA) \cite{lakhina2004diagnosing}, Generalized Likelihood Ratio Test \cite{deshpande2009online} and t-test \cite{ganiz2006detection}. In recent years, Andrian et al. \cite{putina2020online} implemented DenStream, an anomaly detection engine based on clustering techniques, and applied it to a large testbed consisting of dozens of routers. Tal et al. \cite{shapira2022ap2vec} proposed the AP2Vec method, which embeds both ASes and IP address prefixes into feature vectors for BGP hijacking detection. BGP anomaly detection still has a lot of room for research in unsupervised anomaly detection techniques.

\subsection{Time Series Unsupervised Anomaly Detection}
Network traffic anomalous events can be viewed as anomaly detection problems under multivariate time series and the design of unsupervised methods for this problem is highly challenging and widely studied in many fields.

\textbf{Classical unsupervised methods} are widely used for time series data anomaly detection, distance-based (K-Nearest Neighbor, KNN \cite{hautamaki2004outlier}) methods modeled the local behavior around each data point, density-based (Local Outlier Factor, LOF \cite{breunig2000lof}) methods discovered outliers by judging the density around the object domain, clustering-based (Gaussian Mixture Model, GMM \cite{bahrololum2008anomaly}) methods grouped similar data points, and projection-based (PCA \cite{lakhina2004characterization}) and classification-based (One-class SVM, OCSVM \cite{erfani2016high}) methods performed spatial separation of the original data by a linear transformation to achieve anomaly detection.

\textbf{Deep learning-based unsupervised methods} have numerous perspectives to model inter-metric dependencies. Convolutional networks have been applied to sequences for decades, and Bai et al. \cite{bai2018empirical} proposed a Temporal Convolutional Network (TCN) that focuses more on the importance of temporal properties. On modeling inter-metric dependencies, Thill et al. \cite{thill2019anomaly} and Ergen et al. \cite{ergen2019unsupervised} used LSTM for anomaly detection of ECG and HTTP data after preprocessing the data with sliding windows for dimensionality enhancement, and designed window-based error correction methods to optimize the results. AE detects anomalies by deviations between encoded and decoded data of time series. Mo et al. \cite{mo2022unsupervised} reconstruct time series using TCN-AE and Hou et al. \cite{hou2019lstm} use LSTM-AE for higher-order feature learning, both of which are groundbreaking in anomaly detection. A relatively new area is the application of GAN \cite{jiang2019gan}. Similarly, Liang et al. \cite{liang2021robust} proposed a new multi-timescale deep convolutional generative adversarial network (MTS-DCGAN) framework for industrial time series anomaly detection. Based on the several approaches mentioned above, most of the existing studies focused on enhancement and fusion. DAGMM \cite{zong2018deep} jointly optimized the parameters of the deep autoencoder and the mixture model simultaneously in an end-to-end fashion. MTAD-GAT \cite{zhao2020multivariate} combined windows with GAT to learn complex dependencies of multivariate time series in time and feature dimensions, and jointly predicted and reconstructed models to achieve detection functions. MAD-GAN \cite{li2019mad} took window subsequences with different resolutions as input and captured the temporal correlation of time series distribution using LSTM-RNN. OmniAnomaly \cite{su2019robust} identified anomalies using key techniques such as random variable concatenation, planar normalized flow, and robust representation after segmenting the sequence with a window. The more targeted mining analysis of these techniques improves the performance of each domain further, while also observing the high frequency of windows in both data analysis and model construction. The performance of such methods depends heavily on the length of the time window and therefore requires targeted window lengths to provide satisfactory performance. Some of the methods consider data with different window sizes simultaneously in a multi-scale manner, which brings the problem of data dimension multiplication along with stable performance. The relationship between performance and consumption needs to be considered.

Undoubtedly, the main trends of existing methods are focusing on temporal properties guided by LSTM, and concerned with reconstructive adversarial methods, mainly AE and GAN. At the same time, facing the problems of irregular size and increased resource consumption brought by the window strategies found above, we expect to design adaptive window optimization models to achieve efficient and generalized stable anomaly detection performance.

\section{Methodology}
\begin{figure*}[ht]         
\centering
\setlength{\abovecaptionskip}{-0.2cm}
{
    \begin{minipage}[b]{\linewidth}
        \centering
        \includegraphics[scale=1]{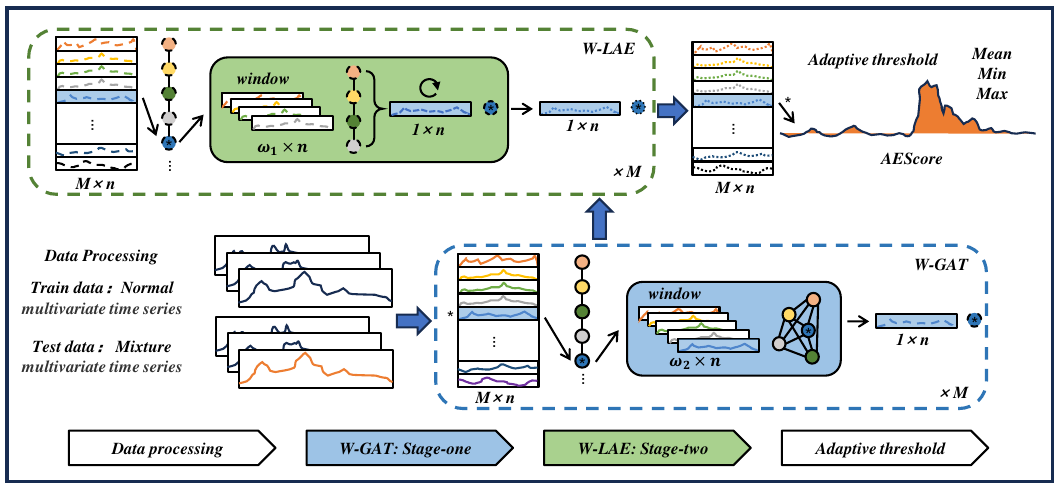}
    \end{minipage}
}
\caption{Model Framework, MAD-MulW. The model has four main components: data processing, stage-one window, stage-two window, and adaptive threshold strategy. Among them, we indicate in detail the learning of sample relationships and retention of the last sample by GAT (blue); and the reconstruction and prediction of samples by LSTM-AE (green). Finally, the sample scores are obtained without changing the data dimensionality to achieve anomaly detection.}
\label{fig: model}
\vspace{-4mm}
\end{figure*}

We propose the unsupervised anomaly detection model MAD-MulW for the current problems of low feature utilization, high model fluctuation, and weak event generalization in the field of BGP anomaly detection. By integrating multi-windows, our model enhances its robustness and generalization ability while maintaining guaranteed detection accuracy. This section provides a comprehensive introduction to MAD-MulW, which contains problem definition, data processing, multi-window structure, and model as a whole, and its overall structure is shown in Figure \ref{fig: model}.

\subsection{Problem Definition}
Our study concentrates on diverse large-scale network security events, specifically traffic monitoring and characterization within a specified range. This paper focuses on features of the communication flow change information, which can be regarded as a multivariate time series of unsupervised anomaly detection. The input data is defined as $\mathcal{T}=\left\{\boldsymbol{x}_{1}, \boldsymbol{x}_{2}, \ldots, \boldsymbol{x}_{M}\right\}, \boldsymbol{x} \in \mathbb{R}^{n}$, $M$ represents the number of time series samples, and $n$ represents the feature dimension corresponding to each time series sample. Therefore the $t$th sample can be represented as $\boldsymbol{x}_{t}=\left\{x_{t}^{1}, x_{t}^{2}, \ldots, x_{t}^{n}\right\}$. It should be noted that the univariate setting $n = 1$ is a special case of multivariate time series \cite{audibert2020usad}.

We define the binary variable $y \in {0, 1}$, where $y_t$ is labeled as 0 if no unusual event occurs at timestamp $t$ and 1 otherwise. To address the problem of unsupervised anomaly detection, we study the data features associated with anomaly detection. The training input $\mathcal{T}$ should consist solely of normal samples, denoted as $y_t = 0$ for $1 \leq t \leq M$. A score of a trained model measures the difference between the invisible sample $\boldsymbol{\hat{x}_{t}}, t>M$ (All subsequent vectors with the symbol $\hat{}$ represent invisible samples.) and $\mathcal{T}$. The normal or abnormal label of the invisible sample are obtained by adaptive threshold strategy.

\subsection{Data Processing}
Standard data processing methods include data cleaning, data normalization, and feature engineering, among others. We process time series datasets in a targeted way to enhance subsequent model learning.

First, we check the data for missing values. If any are found, we fill the vacant part with feature values from the previous time step. Subsequently, we manually clean the columns containing unlearnable features. If the time series' order already captures moment-specific information, the corresponding feature becomes redundant and can be cleaned. Then, since the dataset comes with labels and our model is an unsupervised model, we separate the features and labels. Finally, the dataset is divided into a training set with all normal samples and a testing set with both normal and abnormal samples in a certain proportion.

\subsection{Multi-window Structure}
Time series are characterized by their inherent temporal continuity. On the one hand, this property can be exploited to re-represent time series and construct datasets with enhanced temporal properties. Thus, we propose and design a stage-one window module W-GAT, which employs self-attentive generation to optimize the dataset. On the other hand, we incorporate a stage-two window module W-LAE that replaces individual time points with groups. This is achieved by utilizing the LSTM model in conjunction with the AE model, enabling prediction and anomaly detection in time series data.

\subsubsection{\textbf{Stage-one window W-GAT}} 
\begin{figure}[htb]
\centering
\setlength{\abovecaptionskip}{-0.2cm}
{
    \begin{minipage}[b]{\linewidth}
        \centering
        \includegraphics[scale=01]{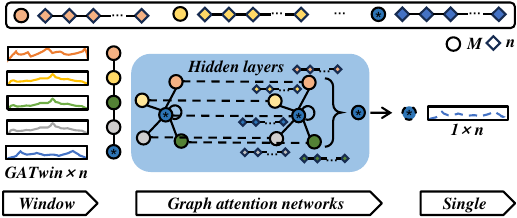}
    \end{minipage}
}
\caption{stage-one window W-GAT.}
\label{fig: GAT}
\end{figure}

Time series representation is a process of reshaping the data, which facilitates the completion of subsequent model tasks and more significantly improves model efficiency. Ding et al. \cite{ding2023mst} used GAT to explicitly model the different relationships in multimodal time series to obtain a better representation of the input data. We designed a stage-one window module W-GAT for reshaping the representation of time series data, which consists of two parts, the first part is a stage-one window design and the second part is a self-attentive model, as shown in Figure \ref{fig: GAT}.

\textbf{Stage-one window group:} The first part is applicable to all datasets. We design a uniform window size parameter $w_1$ and wish to represent a timestamp $\boldsymbol{x_t}$ with the past timestamps $[\boldsymbol{x_{t-w_1+1}},...,\boldsymbol{x_{t-1}},\boldsymbol{x_t]}$. A fully connected network is constructed using graph nodes to represent each timestamp in the window, and weights \textbf{$\alpha$} to indicate the degree of association. The final theoretical representation is

\begin{equation}
    \boldsymbol{x_t} = \sum_{i=0}^{w_1-1}\alpha_{i}\boldsymbol{x_{t-i}}
\end{equation}

\textbf{Graph attention networks:} It is difficult to capture the optimal situation of different datasets by manually adjusting the parameter \textbf{$\alpha$}. So we utilize the self-attention network to correlate the time series and obtain the timestamp attention weights automatically.

The input of stacking graph attention layers is a set of timestamp feature vectors, $[\boldsymbol{x_{t-w_1+1}},...,\boldsymbol{x_{t-1}},\boldsymbol{x_t]}$. Update the current timestamp $\boldsymbol{x_t}$ with the past timestamp to get the final output sample $\boldsymbol{x_{i}^{\prime}}$. The formula for obtaining $\boldsymbol{x_{i}^{\prime}}$ is as follows:

\begin{equation}
    \boldsymbol{x_{i}^{\prime}}=\sigma\left(\sum_{j \in \boldsymbol{N_{i}}} \alpha_{ij} \boldsymbol{W} \boldsymbol{x_{j}}\right)
\end{equation}

where $\boldsymbol{W}$ is the weight matrix, $\sigma(\cdot)$ is the nonlinear activation function , $\boldsymbol{N_{i}}$ represents the set of all neighbors of timestamp $i$, and $\alpha_{ij}$ is attention coefficient in GAT which is calculated as:

\begin{small}
\begin{equation}
   \alpha_{i j}=\frac{\exp \left(\operatorname{\emph{LR}}\left(\mathrm{\boldsymbol{{a}^{T}}}\left[\boldsymbol{W}\boldsymbol{x_{i}} \| \boldsymbol{W} \boldsymbol{x_{j}}\right]\right)\right)}{\sum_{k \in \boldsymbol{N_{i}}} \exp \left(\operatorname{\emph{LR}}\left(\mathrm{\boldsymbol{{a}^{T}}}\left[\boldsymbol{W x_{i}} \| \boldsymbol{W x_{k}}\right]\right)\right)}
\end{equation}

\end{small}where ${(\cdot)}^T$ represents transposition and $||$ is the concatenation operation. $\boldsymbol{a}\in\mathbb{R}^{2n}$ is the weight matrix between the connected layers. The $LR$ (i.e. $LeakyReLu$) function is applied to the output layer of the feedforward neural networks, which resets all negative numbers of the output to 0.2.

$\boldsymbol{x_{i}^{\prime}}$ is a new feature vector for timestamp $i$, which reinforces the role of feature information from past time on current time.

The stage-one window W-GAT plays a role in reshaping the features in the overall model and is the key point of this paper. Unlike previous work that utilized GAT to capture strong and weak correlations at the time level or feature level. This section uses samples as the updating unit and focuses only on the degree of influence of all historical samples on the most recent sample. By using GAT to adaptively update the weights, the stability of the input information is improved by smoothing the current feature fluctuations with historical information. At the same time, it ensures that the dimension of the output data is equal to the dimension of the original data, avoiding the increase in space cost brought by GAT.

\subsubsection{\textbf{Stage-two window W-LAE}}
\begin{figure}[htb]
\centering
\setlength{\abovecaptionskip}{-0.2cm}
{
    \begin{minipage}[b]{\linewidth}
        \centering
        \includegraphics[scale=01]{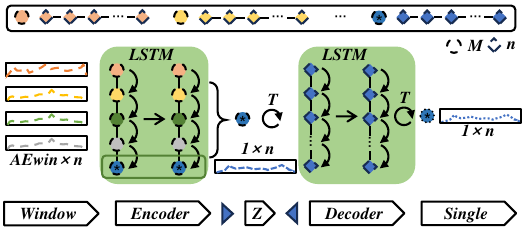}
    \end{minipage}
}
\caption{stage-two window W-LAE.}
\label{fig: LSTM-AE}
\end{figure}

The focus of the stage-two window is on data reconstruction based on time-series prediction. Distinguishing from traditional autoencoders, our approach specifies a window size of historical samples, predicts the sample information at the next moment, and reconstructs the individual timestamps using the property of multi-featured variables, i.e., the size of the encoder's inputs is not equal to the size of the outputs. This design captures the temporal relationship between samples and also increases the variability between different classes of samples to achieve improved detection performance. The structure of stage-two window W-LAE is shown in Fig \ref{fig: LSTM-AE}.

\textbf{Stage-two window group:} The first part of the window group is designed to allow multiple previous timestamps to replace the current timestamp. In this case, we set the stage-two window parameter $w_2$. Then the current time series $\boldsymbol{x_t^{\prime}}$ can be represented as $[\boldsymbol{x_{t-w2}^{\prime}},\dots,\boldsymbol{x_{t-2}^{\prime}},\boldsymbol{x_{t-1}^{\prime}}]$. Note that here the window group represents a certain timestamp but does not contain the time series at that time so that the purpose of predicting the unknown time series from known previous data can be achieved.

\textbf{L-AutoEncoder:} AE is the focus of the reconstructed model, and combining LSTM with AE for prediction function. AE mainly consists of two parts, encoder, and decoder, as shown in Figure \ref{fig: LSTM-AE}. The role of the encoder is to convert the input $x_t^{\prime}$ to the hidden variable $\boldsymbol{z_t}$ by the neural network learning to a specific mapping relation. Usually, it is a nonlinear mapping function:
\begin{equation}
    \boldsymbol{z_{t}}=f\left(\boldsymbol{W [\boldsymbol{x_{t-w2}^{\prime}},\dots,\boldsymbol{x_{t-2}^{\prime}},\boldsymbol{x_{t-1}^{\prime}}]+b}\right)
\end{equation}
where $\boldsymbol{W}$ is the weight between input $\boldsymbol{x_t^{\prime}}$ and hidden variable $\boldsymbol{z_t}$ and $\boldsymbol{b}$ is the bias.

The role of the decoder is the process of reconstructing and reducing the hidden variable $\boldsymbol{y_t}$ to the initial dimension $\boldsymbol{x_{t}^{\prime}}$. The mapping process can be expressed as follows:
\begin{equation}
    \boldsymbol{y_{t}}=f^{\prime}\left(\boldsymbol{W^{\prime} z_{t}+b^{\prime}}\right)
\end{equation}
where $\boldsymbol{W^{\prime}}$ is the weight between hidden variable $\boldsymbol{y_t}$ and output $\boldsymbol{z_{t}}$ and $\boldsymbol{b^{\prime}}$ is the bias. The best state is that the output of the decoder can perfectly or approximately recover the original input, i.e., $\boldsymbol{y_{t} \approx x_t^{\prime}}$.

The main feature of the W-LAE module in this part is that the neural network used for the mapping process is fixed as LSTM. The principle of training W-LAE is to minimize the reconstruction error. On the one hand, it retains the strong generalization of the reconstruction-based approach under unsupervised tasks, and on the other hand, it also preserves the information on the relationship between features in the process of encoding and decoding. Meanwhile, the hidden variable $z_t$ can be regarded as an important feature extracted from the input data sufficiently for subsequent analysis. All these factors guarantee the excellent performance of W-LAE in the unsupervised anomaly detection task, which will be verified in the subsequent experiments.

\subsection{Our Model}
Our unsupervised anomaly detection model is mainly divided into a training part and a testing part, and these parts of the model are made to unite into a whole by designing suitable loss functions and score functions by ourselves.

The original dataset has been cleaned up after data processing. We completed the data reshaping through a stage-one window W-GAT, which step changed the data feature values in the time dimension to reduce noise in continuous samples. After that, the prediction task is implemented by the W-LAE module with data grouping and data reconstruction. Finally, we compare the prediction time series $\mathcal{\hat{T}}=\{\boldsymbol{\hat{x}_1}, \boldsymbol{\hat{x}_2}, \dots, \boldsymbol{\hat{x}_N\}}$ with the original data $\mathcal{T}=\{\boldsymbol{x_1}, \boldsymbol{x_2}, \dots, \boldsymbol{x_N}\}$ for the reconstruction error calculation as follows:

\begin{equation}
    Error = \frac{\sum_{i=1}^n|\boldsymbol{\hat{x}_i}-\boldsymbol{x_i}|}{n}
\end{equation}

The reconstruction error is returned as a loss function during the training process, with the aim of making the prediction result similar to the actual timestamp. The optimal loss function is as follows:

\begin{equation}
    \min Loss=\frac{\sum_{i=1}^n|\boldsymbol{\hat{x}_i}-\boldsymbol{x_i}|}{n}
\end{equation}

During testing, the reconstruction error will be used as $AEScore$ combined with dynamic thresholds for anomaly detection. The dynamic thresholds are selected as the minimum value $Min$, maximum value $Max$, and mean value $Mean$ in $AEScore$, and the $SlideStep$ of the threshold and the $SlideRange$ of the threshold is calculated. The calculation steps are as follows:

\begin{equation}
    SildeStep = \frac{|Max-Min|}{100}
\end{equation}
\begin{equation}
    SlideRange = Mean \pm 25 \times SlideStep
\end{equation}

Following multiple rounds of training and optimization based on the complete process described above, the model performance can be validated on a test set.

\section{Experimental Setup}

\subsection{Datasets}
The datasets used in this thesis are derived from several classes of classical BGP anomalous events \cite{fonseca2019bgp}. We summarize the information of each dataset in Table \ref{tab: datasets}, and its detailed description is shown below.

\begin{table}[htbp]
\caption{BGP Anomaly Event Datasets}
\renewcommand\arraystretch{1.5}
\begin{tabular*}{\hsize}{@{}@{\extracolsep{\fill}}c||cccc@{}}
\bottomrule[0.5mm]
Dataset & Total & \makecell[c]{Anomaly\\(Rate)} & Features & Time or Field  \\    \hline
Code Red II & 7136      & \makecell[c]{472\\(6.61\%)}    & 48      & 2001.07.17-2001.07.21   \\
Nimda       & 10336     & \makecell[c]{353\\(34.2\%)}    & 48      & 2001.09.15-2001.09.23 \\
Slammer     & 7200      & \makecell[c]{1130\\(15.69\%)}    & 48      & 2003.01.23-2003.01.27 \\
Moscow      & 7200      & \makecell[c]{171\\(2.38\%)}    & 48      & 2005.05.23-2005.05.27\\
Malaysian   & 7200      & \makecell[c]{185\\(2.57\%)}    & 48      & 2015.06.10-2015.06.14 \\
\toprule[0.5mm]
\end{tabular*}
\label{tab: datasets}
\end{table}

\textbf{Code Red II}: On July 19, 2001, the Code Red II worm started its spreading across the global network. Since the beginning of the anomaly, it was observed an exponentially growing eight-fold increase in the BGP advertisement rate, over a period of about eight hours.

\textbf{Nimda}: On September 18, 2001, as the Nimda worm started its propagation, Over a period of roughly two hours, the rrc00 collector perceived BGP advertisement rates exponentially ramped up by a factor of 25. The advertisement rate then decayed gradually over several days.

\textbf{Slammer}: The Slammer worm was released on Jan 25, 2003, 5:31 UTC, infected at least 75,000 hosts in just over 30 minutes and it was reported that a number of critical AS-AS peering links were operating above critical load thresholds during the attack period.

\textbf{Moscow Blackout}: The Moscow Power Blackout occurred on May 25, 2005, and lasted several hours. The effect was apparent at the RIS remote route collector in Vienna (rrc05) through a surge in announcement messages arriving from peer AS 12793.

\textbf{Malaysian Telecom Leak}: Malaysian Telecom (AS 4788) leaked one-third of the IP prefixes in its global routing table to backbone provider Level 3 (AS 3549) from June 12, 2015. It left the company inundated with data, resulting in severe packet loss and performance degradation.

\subsection{Comparative Methods}
In order to ensure comprehensive coverage of the comparison methods, we carefully selected 11 techniques that encompass a wide range of unsupervised approaches, including both classical and the latest methods. Among them, the \textbf{classical methods} include KNN, LOF, Connectivity-Based Outlier Factor (COF), Cluster-based Local Outlier Factor (CBLOF), Histogram-based Outlier Score (HBOS), Isolation Forest (iForest), OCSVM, PCA. The \textbf{latest methods} include DAGMM, MTAD-GAT, MAD-GAN, and OmniAnomaly, and these methods have been widely approved and used in time series anomaly detection work in recent years.

\subsection{Evaluation Metrics}
We view the BGP anomaly detection as a classification problem, thus choosing $Accuracy$, $Precision$, $Recall$, and $F1$ as the evaluation metrics. The dataset of real anomalous events exhibits a significant class imbalance. Among them, $F1$ is a weighted average of model precision and recall that can be used to measure the precision of unbalanced data. To facilitate comparison, we assign the abnormal samples as the positive class and the normal samples as the negative class. This allows classification into  True Positive (TP), False Positive (FP), True Negative (TN), or False Negative (FN).

According to the above four situations, we can calculate each performance metric as follows:
\begin{small}
\begin{align}
    \text {\emph{Accuracy}}&=\frac{T P+T N}{T P+T N+F P+F N}\\
    \text {\emph{Precision}}&=\frac{T P}{T P+F P}\\
    \text {\emph{Recall}}&=\frac{T P}{T P+F N}\\
    F 1&=\frac{2 \times \text {\emph{Precision}} \times \text {\emph{Recall}}}{\text {\emph{Precision}}+\text {\emph{Recall}}}
\end{align}
\end{small}

\subsection{Model Parameters}
The performance of a model is sensitive to variable parameters. We split the normal data into a training set and a testing set with a ratio of 7:3, the training set consists of normal data and the test set contains normal and abnormal data. The training parameters are chosen as follows: learning rate is $1e-2$, the epoch is $10$, the W-GAT window is $15$, and W-LAE is $11$. Part of the parameter setting of the model is adjusted according to the aggregation of the concrete experiment. This selection can make the model achieve an excellent and stable result on many datasets. Our model and all experiments are implemented on \emph{Python}, relying on the \emph{PyTorch} framework, the \emph{sklearn} library, and other related libraries and functions.

\begin{table*}[htbp]
\caption{Experiment results}
\renewcommand\arraystretch{1.5}
\begin{tabular*}{\hsize}{@{}@{\extracolsep{\fill}}c||c|cccccccccccc@{}}
\bottomrule[0.5mm]
\multicolumn{1}{c}{\multirow{2}{*}{Dataset}} & \multicolumn{1}{c}{\multirow{2}{*}{Model}} & \multicolumn{7}{c}{Classical Model} & \multicolumn{4}{c}{Latest Model} & \multicolumn{1}{c}{\multirow{2}{*}{\textbf{Ours}}} \\ \cmidrule(r){3-9} \cmidrule(r){10-13}
\multicolumn{1}{c}{} & \multicolumn{1}{c}{}  & KNN   & LOF   & CBLOF & HBOS  & iForest              & OCSVM & PCA   & MTAD                 & MAD                  & DAGMM                 & Omni  & \multicolumn{1}{c}{}               \\  \hline
& Acc.    & 85.29 & 85.79 & 85.03 & 85.31 & 87.22                & 84.87 & 85.85 & 88.96                & 76.62                & 94.96  & 90.65 & \underline{\textbf{97.82}}                \\
& Pre.    & 51.62 & 53.02 & 50.92 & 51.69 & 56.99                & 50.45 & 53.17 & 66.11                & 64.96                & 65.57                 & 65.81 & \underline{\textbf{96.54}} \\
& Rec. & 52.36 & 54.40 & 51.34 & 52.47 & 60.18                & 50.65 & 54.63 & 81.33 & 64.04                & 50.53                 & 72.79 & \underline{\textbf{96.13}}                \\
\multirow{-4}{*}{Code Red II} & F1     & 51.72 & 53.38 & 50.89 & 51.81 & 57.07                & 50.34 & 53.56 & 70.23                & 64.46                & 57.14                 & 68.43 & \underline{\textbf{96.33}} \\  \hline
& Acc.    & 65.78 & 65.35 & 64.48 & 66.75 & 70.69                & 64.43 & 67.37 & 69.75                & 72.72 & 67.41                 & 68.09 & \underline{\textbf{86.76}}                \\
& Pre.    & 58.72 & 57.54 & 55.12 & 61.41 & 72.37 & 54.96 & 63.13 & 66.28                & 65.48                & 69.41                    & 66.12 & \underline{\textbf{86.21}}                \\
& Rec. & 53.49 & 53.02 & 52.05 & 54.56 & 58.95                & 51.99 & 55.25 & 66.06                & 70.66 & 8.35                  & 65.34 & \underline{\textbf{85.04}}                \\
\multirow{-4}{*}{Nimda}     & F1     & 50.31 & 49.69 & 48.43 & 51.72 & 57.45                & 48.34 & 52.61 & 66.17                & 66.28 & 14.90                 & 65.60 & \underline{\textbf{85.55}}                \\  \hline
& Acc.    & 83.06 & 78.50 & 82.14 & 84.36 & 87.47 & 82.03 & 83.78 & 83.63                & 77.32                & 83.19                 & 80.20 & \underline{\textbf{98.19}}                \\
& Pre.    & 65.59 & 52.93 & 63.04 & 69.21 & 77.85                & 62.73 & 67.59 & 69.54                & 70.20                & 38.24                 & 69.43 & \underline{\textbf{98.12}} \\
& Rec. & 60.60 & 51.99 & 58.87 & 63.07 & 68.95                & 58.66 & 61.97 & 70.79                & 76.58                & 11.50                 & 79.97 & \underline{\textbf{97.93}} \\
\multirow{-4}{*}{Slammer}   & F1     & 62.17 & 51.99 & 60.12 & 65.08 & 72.03                & 59.87 & 63.78 & 70.07                & 71.65                & 17.69                 & 71.72 & \underline{\textbf{98.02}} \\  \hline
& Acc.    & 91.99 & 88.46 & 91.90 & 91.99 & 92.12                & 91.74 & 91.99 & 98.24 & 73.38                & 97.92                 & 98.06 & \underline{\textbf{99.36}}                \\
& Pre.    & 60.79 & 51.00 & 60.56 & 60.79 & 61.18                & 60.10 & 60.79 & 80.92                & 66.16                & 53.85                 & 81.13 & \underline{\textbf{99.67}} \\
& Rec. & 91.90 & 53.86 & 91.00 & 91.90 & 93.40                & 89.21 & 91.90 & 80.92                & 71.55                & \underline{\textbf{100.00}} & 71.13 & 91.03                \\
\multirow{-4}{*}{Moscow}                 & F1     & 65.49 & 50.29 & 65.13 & 65.49 & 66.08                & 64.41 & 65.49 & 80.92                & 67.06                & 70.00                 & 75.11 & \underline{\textbf{94.90}} \\  \hline
& Acc.    & 91.71 & 89.35 & 91.51 & 91.51 & 92.26                & 90.96 & 91.99 & 97.13 & 83.52                & 96.94                 & 95.97 & \underline{\textbf{99.26}}                \\
& Pre.    & 60.46 & 53.90 & 59.92 & 59.92 & 62.00                & 58.37 & 61.23 & 70.79                & 76.80                & 45.45                 & 66.41 & \underline{\textbf{99.62}} \\
& Rec. & 87.60 & 64.01 & 85.65 & 85.65 & 93.14                & 80.10 & 90.36 & 69.13                & 70.17                & \underline{\textbf{94.59}}  & 75.83 & 90.24                \\
\multirow{-4}{*}{Malaysian}                & F1     & 64.80 & 54.78 & 63.98 & 63.98 & 67.16                & 61.62 & 65.98 & 70.02                & 72.48                & 61.40                 & 69.94 & \underline{\textbf{94.40}} \\  \hline 
& Acc.    & 83.57 & 81.49 & 83.01 & 83.98 & 85.95                & 82.81 & 84.20 & 87.54                & 76.71                & 88.08  & 86.59 & \underline{\textbf{96.28}}                \\
& Pre.    & 59.44 & 53.68 & 57.91 & 60.60 & 66.08                & 57.32 & 61.18 & 70.73                & 68.72                & 54.50                 & 69.78 & \underline{\textbf{96.03}} \\
& Rec. & 69.19 & 55.46 & 67.78 & 69.53 & 74.92                & 66.12 & 70.82 & 73.65                & 70.60                & 52.99                 & 73.01 & \underline{\textbf{92.07}} \\
\multirow{-4}{*}{Average}                              & F1     & 58.90 & 52.03 & 57.71 & 59.62 & 63.96                & 56.92 & 60.28 & 71.48                & 68.39                & 44.23                 & 70.16 & \underline{\textbf{93.84}}  \\
\toprule[0.5mm]
\end{tabular*}
\label{table: result}
\end{table*}

\section{Experiments and Results}

This section focuses on evaluating the performance of the model in unsupervised anomaly detection and assessing the effectiveness of the stage-two window design for each component of the structure. We then discuss the rationality of each of the three hyperparameter designs separately.

\subsection{Basic Experiment}
In this section, we compare the performance differences of numerous unsupervised anomaly detection methods mainly on several BGP anomaly traffic datasets, and the results are shown in Table \ref{table: result}. Under each metric, the optimal results are bolded and underlined. Due to the small proportion of anomaly samples in datasets, detection becomes significantly more challenging. Consequently, the F1 metric provides a more comprehensive evaluation of the model's performance in both normal and anomalous sample identification. From the experimental results, we can clearly see that the latest methods generally outperform the classical detection models, mainly because the classical methods focus more on the partial attributes of events. Although more efficient, they are not as comprehensive as the latter in terms of detection performance after all. In addition, deep learning-based methods can capture deeper feature information and relationship information, which makes their models show better performance. We calculate the average of each model metric at the end of Table \ref{table: result}, and the results clearly show that our model improves by 20\% on the most critical F1 metric, and generally above 90\% on the other three metrics. Such substantial improvement over several datasets reflects the high performance and strong stability of our model, which is also attributed to the adaptive and sufficient noise reduction of the fluctuation information by the W-GAT and the capture of the temporal correlation property by W-LAE. The model will prove its excellent performance in the subsequent parametric analysis.

\begin{figure*}[ht]
\centering
\setlength{\abovecaptionskip}{-0.2cm}
{
    \begin{minipage}[b]{\linewidth}
        \centering
        \includegraphics[scale=0.65]{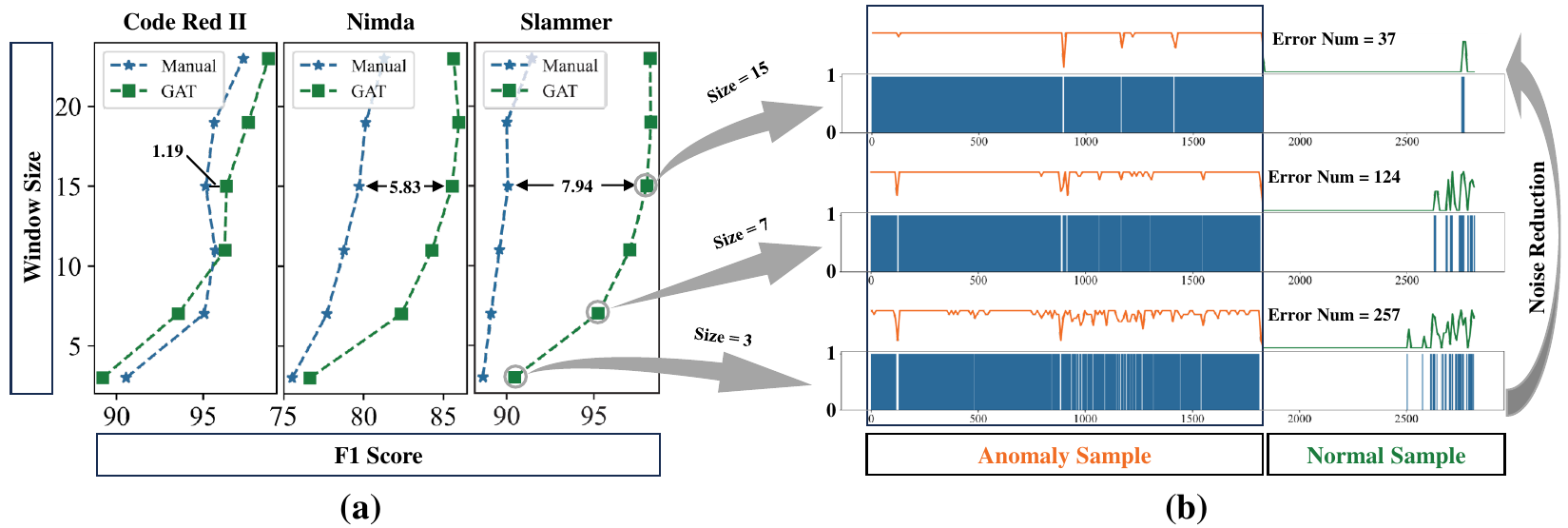}
    \end{minipage}
}
\caption{W-GAT module validation. (a) Performance differences between the manual and adaptive windows; (b) The predicted labels of the test samples under different window sizes.}
\label{fig: window_analysis}
\end{figure*}

\subsection{Window validity}
We have two window parameters, where the W-GAT window is to enhance the temporal correlation and reconstruct features, and the W-LAE window is to make the model achieve the prediction role. Here we verify the effectiveness of the W-GAT window by setting three sets of experiments: (1) no window, (2) manual window, and (3) GAT adaptive window. When the window size is 15, the results of the three datasets are shown in Table \ref{table: window_size}. The findings indicate that the manual window results in a 4\%-8\% increase in the F1 score compared to the absence of a window. Furthermore, the GAT-based adaptive window exhibits a further 1\%-8\% improvement in the F1 score. The current results already demonstrate the positive impact of the window on data enhancement and the significant improvement in detection performance achieved by the adaptive window.

\begin{table}[htbp]
\vspace{-3mm}
\caption{Window experiment results (Window Size = 15)}
\renewcommand\arraystretch{1.5}
\begin{tabular}{ccccc}
\bottomrule[0.5mm]
\multicolumn{1}{c}{\multirow{2}{*}{Description}} & \multicolumn{1}{c}{\multirow{2}{*}{Window Size}} & \multicolumn{3}{c}{Dataset}   \\ 
\cline{3-5} 
\multicolumn{1}{c}{}     & \multicolumn{1}{c}{}     & Code Red II & Nimda & Slammer \\ \hline
No Window                & -                        & 86.31       & 73.46 & 86.32   \\
Manual Window            & 15                       & 95.14       & 79.72 & 90.08   \\
Adaptive Window          & 15                       & \textbf{96.33}  & \textbf{85.55} & \textbf{98.02}   \\
\toprule[0.5mm]
\end{tabular}
\label{table: window_size}
\end{table}

\begin{figure*}[ht]         
\centering
\setlength{\abovecaptionskip}{-0.2cm}
{
    \begin{minipage}[b]{\linewidth}
        \centering
        \includegraphics[scale=0.55]{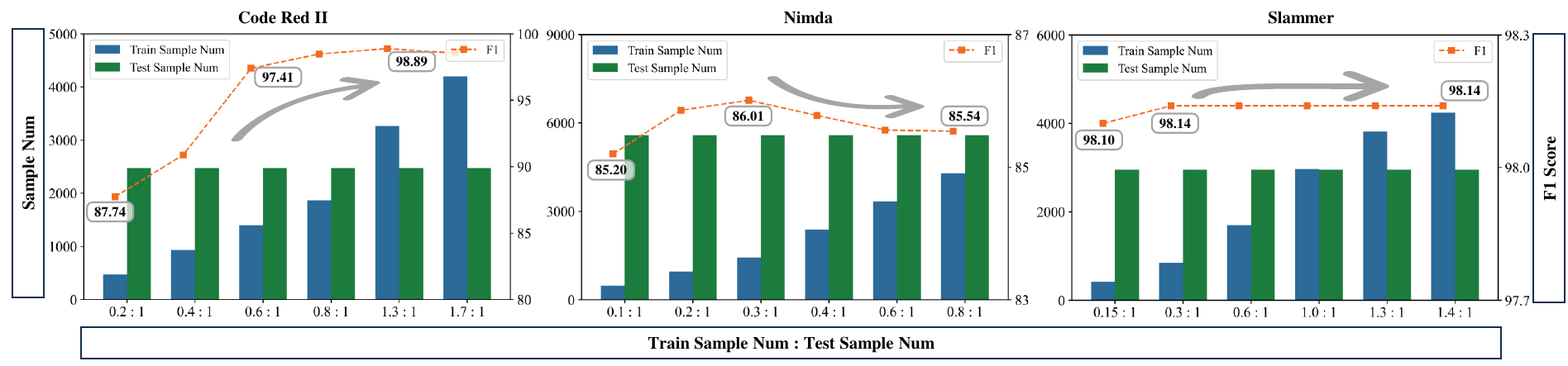}
    \end{minipage}
}
\caption{The variation of the results with different sample ratios.}
\label{fig: ratio}
\vspace{-3mm}
\end{figure*}

However, is it a coincidence that the current window size makes excellent results? To answer this question, we obtain the experimental results in Figure \ref{fig: window_analysis}(a). It is obvious from the three figures that the results of manual and adaptive are similar for small windows in the early stage, but as the window increases, the adaptive window shows a significant improvement in performance, even up to an 8\% difference on the Slammer dataset. The optimal manual window value is not fixed for different datasets, and the impact of the window is inconsistent, showing a tendency to fluctuate upward with the window size. In contrast, the GAT adaptive window can determine the importance of the sample by the parameter value instead of fixing it to 0 or 1. This adaptive nature allows the F1 score to show a steady upward trend as the window size increases, ensuring that each dataset achieves near-optimal results in the GAT window.

A potential explanation is that the adaptive window of GAT simulates a manual window, using the connection weights between the time series within the window to simulate the window size, ultimately achieving a near-optimal result. In particular, the performance improvement of the window from presence to absence is the largest, and the explanation is also related to the presence of noise in the time series itself, as it is a non-ideal curve. The addition of the window effectively smoothes out the noise, eliminates outliers, and ensures overall smoothness. As shown in Figure \ref{fig: window_analysis}(b), the fluctuations of the samples due to misclassification are decreasing significantly with the increase of the window size. Additionally, we artificially assign weights to the manual windows to simulate the adaptive windows of GAT, aiming to assess the impact of weights as opposed to fixed-value windows.

\subsection{Parameter Validity}
In the course of our experiments, three questions arose as follows:
\begin{itemize}
\item[$\bullet$] \textbf{Q1:} What is the effect of the number of training samples on the results?
\item[$\bullet$] \textbf{Q2:} How does the threshold value affect the results?
\item[$\bullet$] \textbf{Q3:} What are the stage-two window parameters set to?
\end{itemize}

\subsubsection{The number of training samples}
The first question is due to the excellent performance and reasonable design of our model in the basic experiments described above, which makes us look forward to further investigating whether the MAD-MulW model can achieve good detection performance with only a small number of training samples. So there was a fixed number of test samples and experiments were conducted for different numbers of training samples, and the experimental results are shown in Figure \ref{fig: ratio}. The horizontal coordinates in the figure are the ratio of the number of train sets to the number of test sets, and the vertical coordinates correspond to the different metric values. The overall trend is more intuitive, and the detection performance improves as the number of train samples increases. It is worth noting that our model generally achieves optimal results using a train set with less than the number of test samples, which again demonstrates the stable and generalized detection performance of our model. The small amount of train data and high detection results indicate that the model requires low learning costs and fast learning. Notably, near-optimal performance is attained even with sparse samples, thereby reducing the restrictiveness of the anomaly detection task and enhancing its applicability in diverse situations.

\begin{figure}[ht]         
\centering
\setlength{\abovecaptionskip}{-0.2cm}
{
    \begin{minipage}[b]{\linewidth}
        \centering
        \includegraphics[scale=0.6]{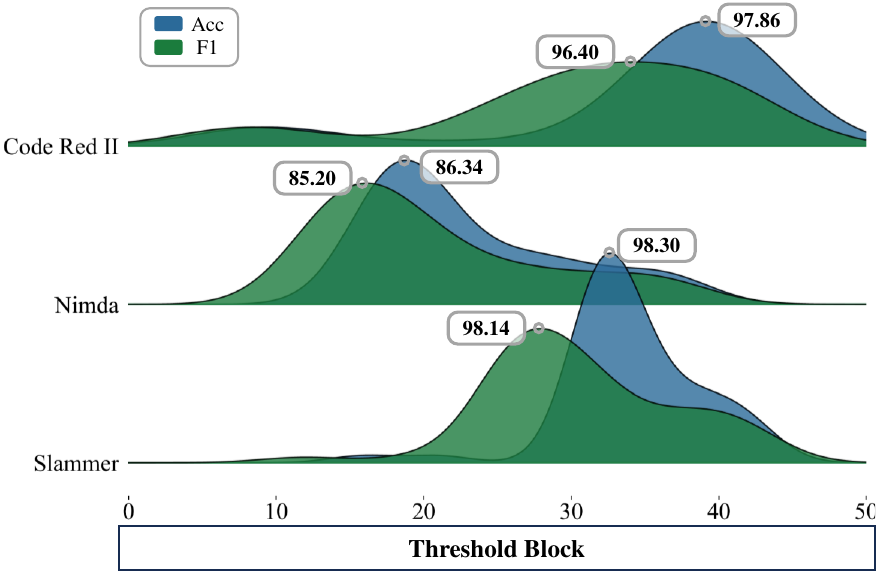}
    \end{minipage}
}
\caption{The variation of the results with different thresholds.}
\label{fig: threshold}
\end{figure}

\subsubsection{Adaptive threshold strategy}
The second question focuses on the study of the adaptive threshold strategy. Figure \ref{fig: threshold} shows the trend of the effect of different thresholds (horizontal coordinates) on the F1 score (vertical coordinates) of the three datasets. Both too-small and too-large thresholds lead to a large number of misclassification cases, while the range of better thresholds is narrower and the selection of thresholds varies on different datasets. Under our adaptive threshold strategy, the divided thresholds do not go through the lowest classification results at the lowest and highest thresholds. Since the change of threshold value has little effect on the division at smoothing, that is the model can achieve good experimental results within a certain range of threshold values. The adaptive strategy clarifies the importance of threshold selection and also shows that achieving dynamic thresholds is a problem that needs to be studied in depth in subsequent work.

\subsubsection{Manual window design}
\begin{figure}[ht]         
\centering
\setlength{\abovecaptionskip}{-0.2cm}
{
    \begin{minipage}[b]{\linewidth}
        \centering
        \includegraphics[scale=0.5]{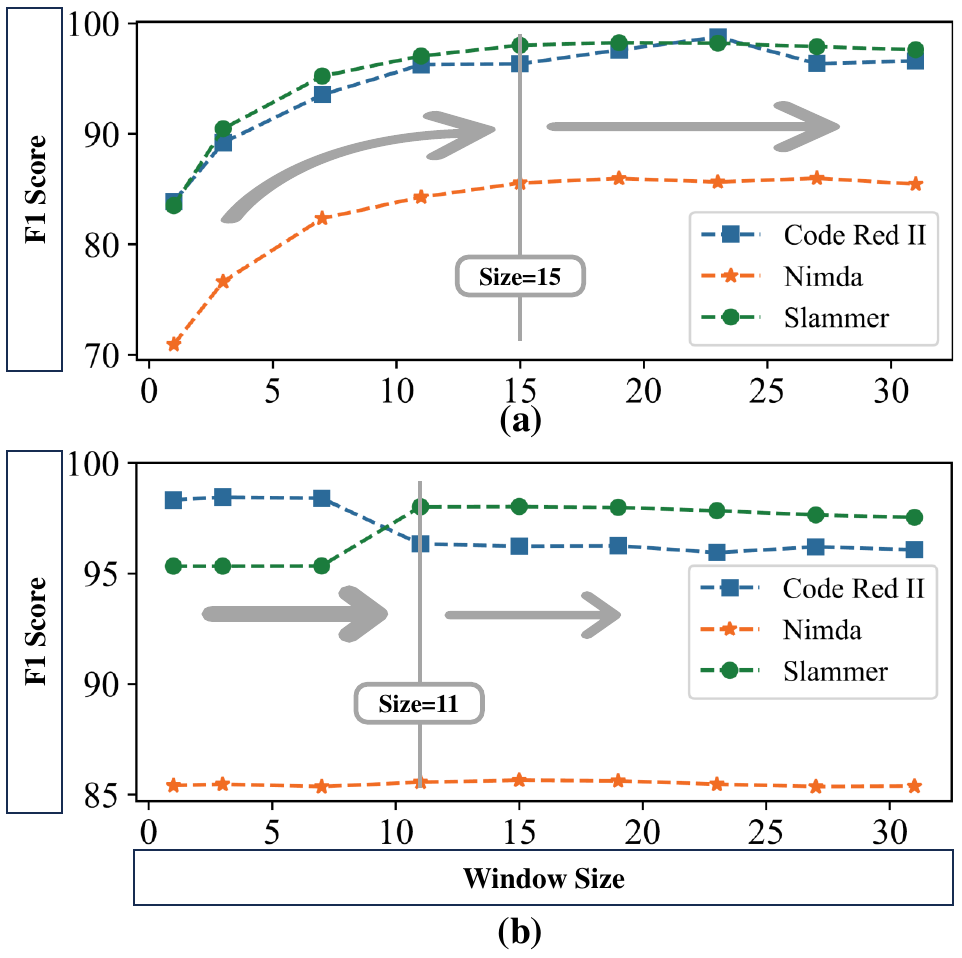}
    \end{minipage}
}
\caption{The variation of the results with different window sizes.}
\label{fig: window_size}
\end{figure}

The third question is the setting of the stage-two window parameters. The trends of detection performance for different datasets with different window parameters are shown in Figure \ref{fig: window_size}.

Upon fixing the size of the AE window, we observed a consistent trend in the size of the GAT window across all datasets: a gradual increase followed by stabilization (Figure \ref{fig: window_size}(a)). To ensure stable and outstanding results across multiple datasets while considering the minimum sample size, we fixed the window size at 15 for the W-GAT module.

The AE window serves to achieve prediction, so the size of the AE window directly affects the temporal memorability of the LSTM during computation. the LSTM needs to learn and infer information from features at known moments to unknown moments, while the long- and short-term dependence of the samples also determines the amount of information required. Our experimental results found (Figure \ref{fig: window_size}(b)) that when the AE window increases, the different datasets do not show a clear and uniform trend. However, when the window is at 11, a better state that can be achieved by considering short-term temporal information.

So far, we have demonstrated the stability and generalization of our model design, as well as the stringency of the parameter design through several experiments.

\section{Conclusion}
In this paper, we innovatively propose a stage-two window model for multivariate time series anomaly detection in order to achieve the detection and analysis of large-scale network anomalous events. The model employs a stage-one W-GAT window for feature reconstruction and a stage-two W-LAE window for time series prediction. Our proposed method outperforms baseline methods and advanced models on the BGP dataset. Furthermore, our model is characterized by low training costs, and fast learning, and achieves excellent detection results even with limited training samples. Moreover, our method demonstrates applicability to a wide range of datasets, thanks to its automatic reshaping capability and stable prediction performance. Future research is likely to come from two directions. One is the follow-up study of Windows, using more real-world scenarios to validate and update the window design. The second is the model architecture. The improvement and modification of the AE model is the core problem of the final anomaly detection, and we will incorporate more learning methods, such as comparative learning, migration learning, etc. The ultimate goal is to have efficient anomaly detection for multiple fields.

\section*{Acknowledgments}
This work was supported in part by the Key R\&D Program of Zhejiang under Grant 2022C01018, by the National Natural Science Foundation of China under Grants 62103374, and U21B2001. All authors are with the Institute of Cyberspace Security, College of Information Engineering, Zhejiang University of Technology, Hangzhou 310023, China. S. Peng, Z. Ruan, and Q. Xuan are also with Binjiang Cyberspace Security Institute of ZJUT, Hangzhou 310056, China.

\bibliographystyle{IEEEtran}
\bibliography{IEEEabrv, references}


 




\vfill

\end{document}